\documentclass[twocolumn,pre,showpacs,aps]{revtex4}
\usepackage{graphicx}
\usepackage{amsmath,amssymb}
\begin{document}

\title{Noise enhanced spontaneous chaos in semiconductor superlattices at room temperature}

\author{ M. Alvaro, M. Carretero, L. L. Bonilla}

\affiliation{Gregorio Mill\'an Institute for Fluid Dynamics, Nanoscience and Industrial Mathematics, Universidad Carlos III de Madrid,
Avenida de la Universidad 30, 28911 Legan\'es, Spain}

\begin{abstract}
Physical systems exhibiting fast spontaneous chaotic oscillations are used to generate high-quality true random sequences in random number generators. The concept of using fast practical entropy sources to produce true random sequences is crucial to make storage and transfer of data more secure at very high speeds. While the first high-speed devices were chaotic semiconductor lasers, the discovery of spontaneous chaos in semiconductor superlattices at room temperature provides a valuable nanotechnology alternative. Spontaneous chaos was observed in 1996 experiments at temperatures below liquid nitrogen. Here we show spontaneous chaos at room temperature appears in idealized superlattices for voltage ranges where sharp transitions between different oscillation modes occur. Internal and external noises broaden these voltage ranges and enhance the sensitivity to initial conditions in the superlattice snail-shaped chaotic attractor thereby rendering spontaneous chaos more robust.  
\end{abstract}



\maketitle

\renewcommand{\thefootnote}{\arabic{footnote}}
\newcommand{\fin}{\newline \rule{2mm}{2mm}}

\section{Introduction} 
\begin{figure}[htbp]
\begin{center}
\includegraphics[width=8cm]{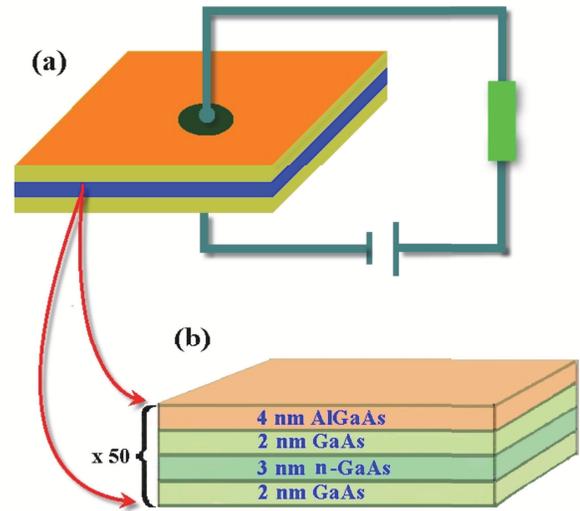}
\end{center}
\caption{Sketch of a mesa-shaped semiconductor superlattice device having 1.2 mm side and 1.5 $\mu$m thickness. (a) dc voltage biased superlattice consisting of two contact regions of about 0.5 $\mu$m width and 50 periods formed by two  semiconductor layers of different bandgaps as depicted in (b). 
} \label{fig1} 
\end{figure}

Spontaneously chaotic semiconductor superlattice (SL) devices (sketched in figures \ref{fig1}(a) and (b)) may become the key to achieve fast true random number generators (RNGs) \cite{uch08,mur08,rei09,kan10,li13}. The latter are crucial to secure fast and safe data storage and transmission \cite{sti95,gal08,nie00}, stochastic modeling \cite{asm07}, and Monte Carlo simulations \cite{bin02}. In ideal spontaneously chaotic systems without noise/uncertainty, with exactly the same initial and boundary conditions, a chaotic RNG would generate the same random number sequence at each repetition of the whole process. Thus ideal deterministic chaotic RNGs correspond to pseudo-random number generators and do not increase information entropy. The randomness in a practical RNG will not be more than that provided by the noise sources therein. 
However, the chaotic system is sensitive to all the changes in the initial, boundary and bulk conditions and possible noises. Thus it may provide more randomness than just the amplification of a simple noise source, because it will generate a cumulative randomness resulting from all these noises, similarly to the situation when we multiply independent noises of zero mean. We show here that internal noise enhances spontaneous chaos at room temperature arising from sharp transitions between different oscillatory modes in SLs. It also induces chaos in voltage intervals close to those corresponding to barely noticeable deterministic chaotic attractors. Thus noise widens the voltage range of spontaneous chaos and it increases the sensitivity of the chaotic attractor to initial conditions thereby improving its speed as a random bit generator. This {\em noise-enhanced chaos}, already demonstrated in simple dynamical systems \cite{cru82,gao98}, is another paradigm of the constructive role of noise similar to stochastic resonance (noise-induced escape from one attractor to another resonates with external force \cite{gam98}) and coherence resonance (noise-induced oscillations in excitable systems \cite{pik97}).

The mechanisms that produce spontaneous chaos in SLs need to be understood to improve design of superlattice-based fast RNGs. Spontaneous chaos at room temperature appears in SLs whose configuration and aluminum content in the barriers are tuned to suppress leakage current through the X valley of the barrier \cite{li13,hua12}. Below liquid nitrogen temperatures, spontaneous chaos was observed in SLs with AlAs barriers in 1996 \cite{zha96}, but leakage currents through their X valley suppressed all oscillations (chaotic or not) for higher temperatures \cite{xia90,zha94}. A generic mechanism to induce spontaneous chaos in bulk semiconductors \cite{can01} and in SLs at ultra low temperatures \cite{ama02,ama03} is based on wave front dynamics. For appropriate values of the injecting contact conductivity (such that the critical current is close to the one at which accumulation and depletion layers move at the same velocity \cite{bon05,can01,ama02}), electric field domains formed by pairs of increasing and decreasing wave fronts are randomly triggered at the injecting contact. The relation of this theoretical mechanism for spontaneous chaos to experiments remains unclear \cite{zha96}. In contrast to this situation, theoretical predictions of {\em driven chaos} under ac+dc voltage bias based on a simple model \cite{bul95} have been observed in experiments \cite{luo98}.
At room temperature, wave fronts are not sharp and therefore it is not obvious that many can coexist simultaneously on a 50-period SL as do the sharp wave fronts seen in numerical simulations at ultra low temperatures \cite{ama02}. We have found a different mechanism that produces spontaneous chaos near voltage regions where sharp transitions between two different oscillation modes occur. Noise enhances this very weak deterministic chaos.

Weakly coupled SLs may act as spatially discrete excitable or oscillatory systems with  local coupling between quantum wells, produced by the inter-well resonant tunneling current and the Poisson equation, and global coupling due to voltage bias \cite{bon05}. The dynamics of charge dipoles and monopoles play a crucial role in both the excitability of a stable stationary state and in generating stable self-sustained oscillations of the current \cite{bon05,BT10}. Both behaviors may appear in different voltage ranges for the same device. 

\section{Model} In our simulations, we have included intrinsic noise in the usual sequential tunneling model of electron transport in a weakly coupled $n$-doped SL and also extrinsic noise due to the voltage source \cite{bon05}
\begin{eqnarray}
&&\varepsilon\frac{dF_i}{dt}+ J_{i\to i+1}+\xi_i(t)=J(t), \label{eq1}\\
&&J_{i\to i+1}=\frac{en_i}{l}v^{(f)}(F_i)-J_{i\to i+1}^-(F_i,n_{i+1},T),
\label{eq2}\\
&& n_i=N_D+\frac{\varepsilon}{e}(F_i-F_{i-1}), \label{eq3}
\end{eqnarray}
\begin{eqnarray}
&& J_{i\to i+1}^-(F_i,n_{i+1},T)=\frac{em^*k_BT}{\pi\hbar^2l}v^{(f)}(F_i)\nonumber\\
&&\quad\times\ln\!\left[1+ e^{-\frac{eF_i l}{k_{B}T}}\!\!\left( e^{\frac{\pi\hbar^{2}n_{i+1}}{m^{*}k_{B}T}} -1\right)\!\right]\!,\label{eq4}\end{eqnarray}
\begin{eqnarray}
&&
\langle\xi_i(t)\xi_j(t')\rangle=\frac{e}{A}\!\left[\frac{ev^{(f)}(F_i)}{l}n_i+J_{i\to i+1}^-(F_i,n_{i+1},T)\right.\nonumber\\
&&\quad\left.+2J_{i\to i+1}^-(F_i,n_{i},T) \right]\!\delta_{ij}\delta(t-t'), \label{eq5}\end{eqnarray}
\begin{eqnarray}
&&J_{0\to 1}=\sigma_0 F_0,\,\, J_{N\to N+1} = \sigma_0\frac{n_N}{N_{D}}F_N, \label{eq6}\\
&&\sum_{i=1}^N F_i=\frac{V+\eta(t)}{l}.\label{eq7}
\end{eqnarray}
Here $i=1,\ldots,N$ ($N=50$ is the number of SL periods \cite{hua12}), and $J_{i\to i+1}$, $\xi_i(t)$, $J(t)$ are the tunneling current density from well $i$ to well $i+1$, the corresponding  fluctuating current density, and the total current density, respectively. The tunneling current density in Eq.~(\ref{eq2}) from well $i$ to $i+1$ depends on the electric field $F_i$ in well $i$ and the two-dimensional electron (2D) densities in the corresponding wells, $n_i$ and $n_{i+1}$. The forward velocity $v^{(f)}(F_i)$ is a function given in \cite{bon05,bon06} with peaks corresponding to three energy levels at 53, 207, and 440 meV calculated by solving a Kronig-Penney model for the SL configuration of References \cite{li13,hua13,hua12}. The level broadenings due to scattering are 2.5, 8 and 24 meV for the three energy levels \cite{bon06}. The equivalent 2D doping density due to the doping of the central part of the quantum well is $N_D=6 \times 10^{10}$ cm$^{-2}$. Also $m^*=(0.063 + 0.083x)m_e=0.1 m_e$ (for $x=0.45$), $-e<0$, $A=s^2$ with $s= 1.2$ mm, $l_b=4$ nm, $l_w=7$ nm, $l=l_b+l_w$, $\varepsilon=l/[\frac{l_w}{\varepsilon_w}+\frac{l_b}{\varepsilon_b}]$, $\varepsilon_b= 10.9 \epsilon_0$, $\varepsilon_w=12.9 \epsilon_0$, $\epsilon_0$, $k_B$, $T$, $V$, and $\sigma_0$ are the effective electron mass, the electron charge, the SL cross section, the side length of a square mesa, the (Al,Ga)As barrier thickness, the GaAs well thickness, the SL period, the SL permittivity, the barrier permittivity, the well permittivity, the dielectric constant of the vacuum, the Boltzmann constant, the lattice temperature, the dc voltage, and the contact conductivity, respectively. The fluctuating currents are independent identically distributed (i.i.d.) zero-mean white noises. Although these noises are i.i.d., the strongly nonlinear character of the system causes the noise sources to influence each other. The first two terms in their correlations in Eq.~(\ref{eq5}) are the SL shot noise \cite{bon04}, and the last term is an effective thermal noise similar to that describing current fluctuations in the bulk Gunn effect \cite{kei87,BT10}. The total current $J(t)$ can be calculated from Eq.~(\ref{eq1}) and the bias condition in Eq.~(\ref{eq7}), thereby providing effective nonlocal equations of motion. $\eta(t)$ is the unavoidable noise of the voltage source characterized as an independent zero-mean Gaussian white noise with 0.28 mV standard deviation. \\

\section{Results} 
\begin{figure}[htbp]
\begin{center}
\includegraphics[width=8cm]{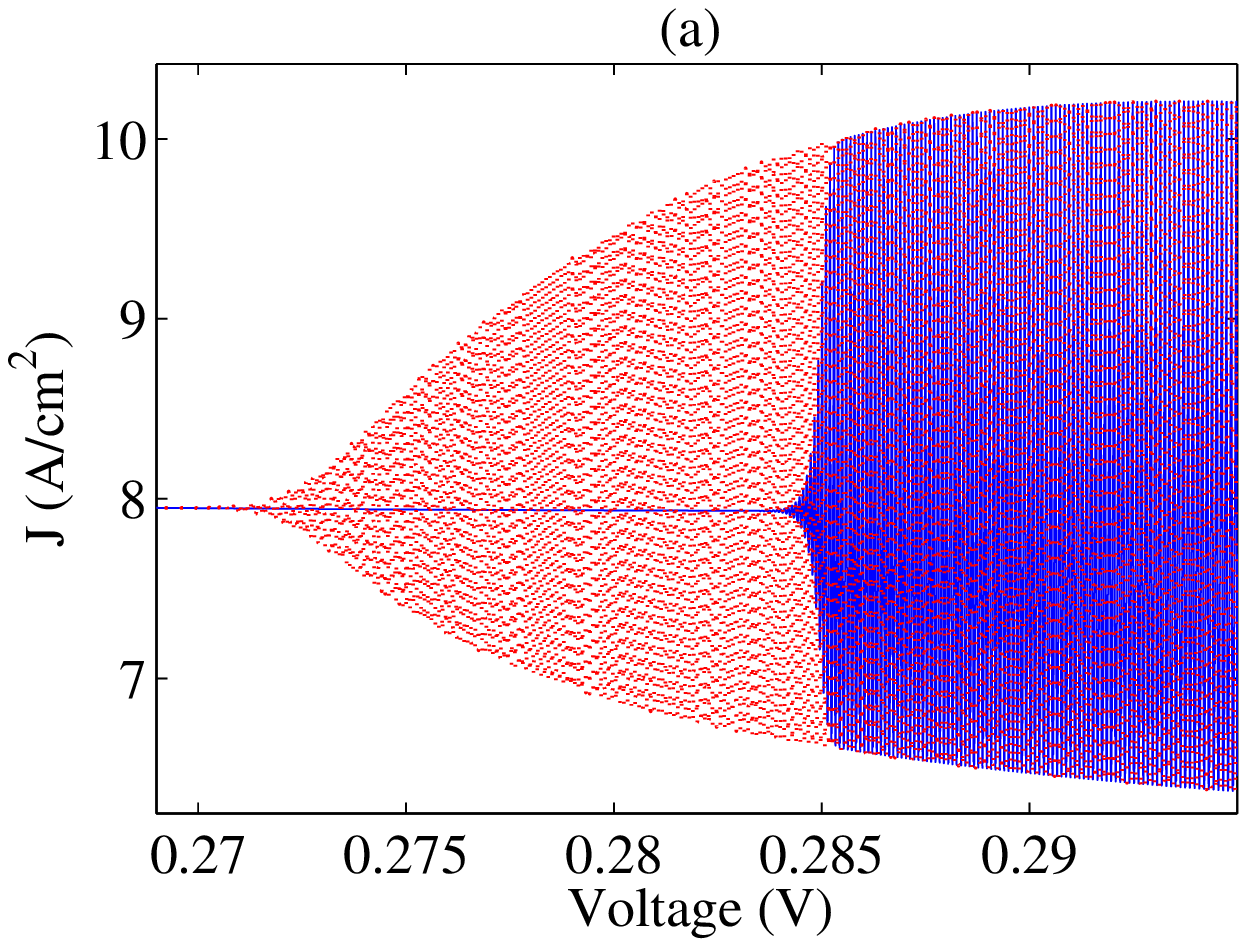}\\
\includegraphics[height=6.5cm,width=8cm]{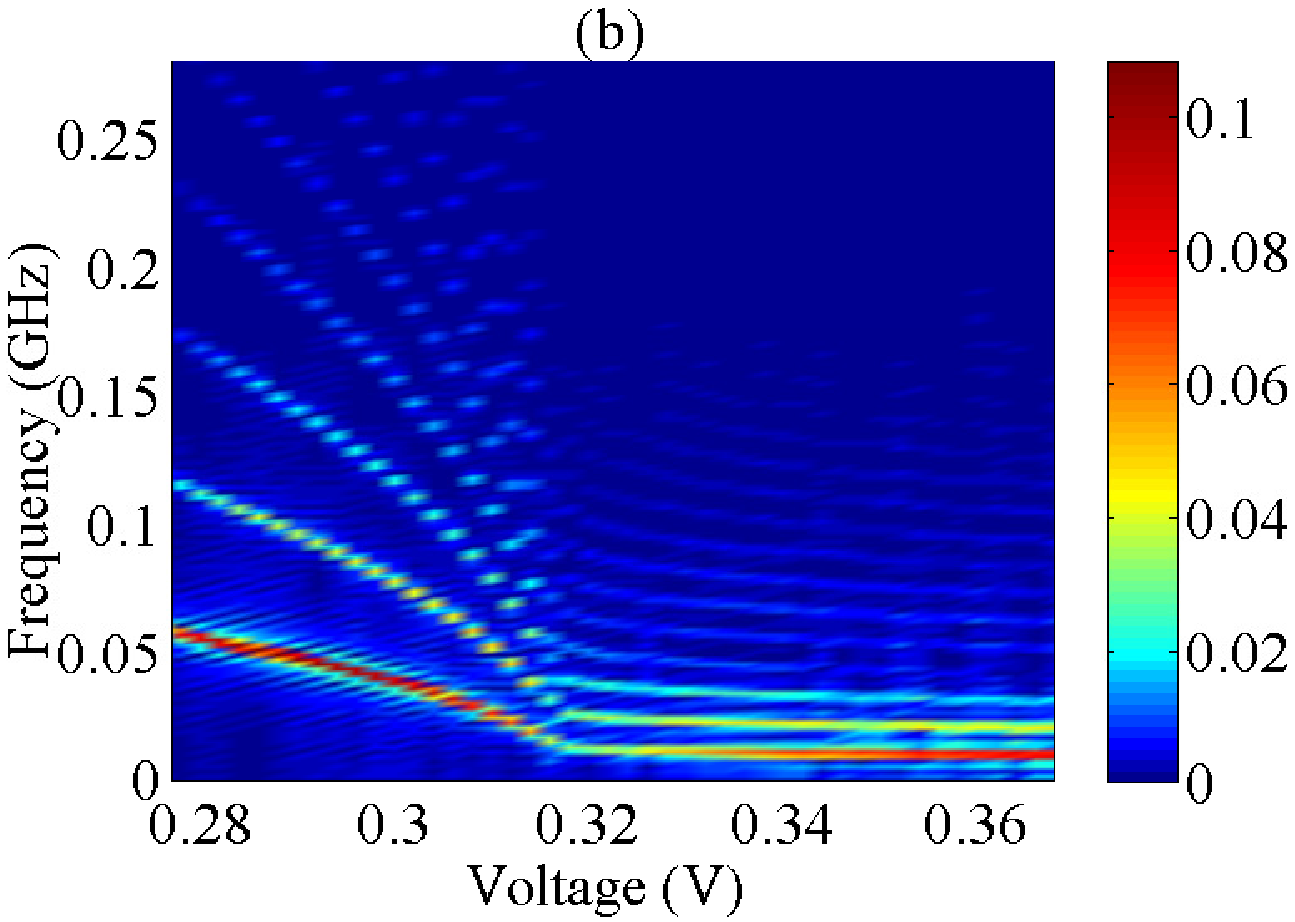}\\
\includegraphics[height=7cm,width=8cm]{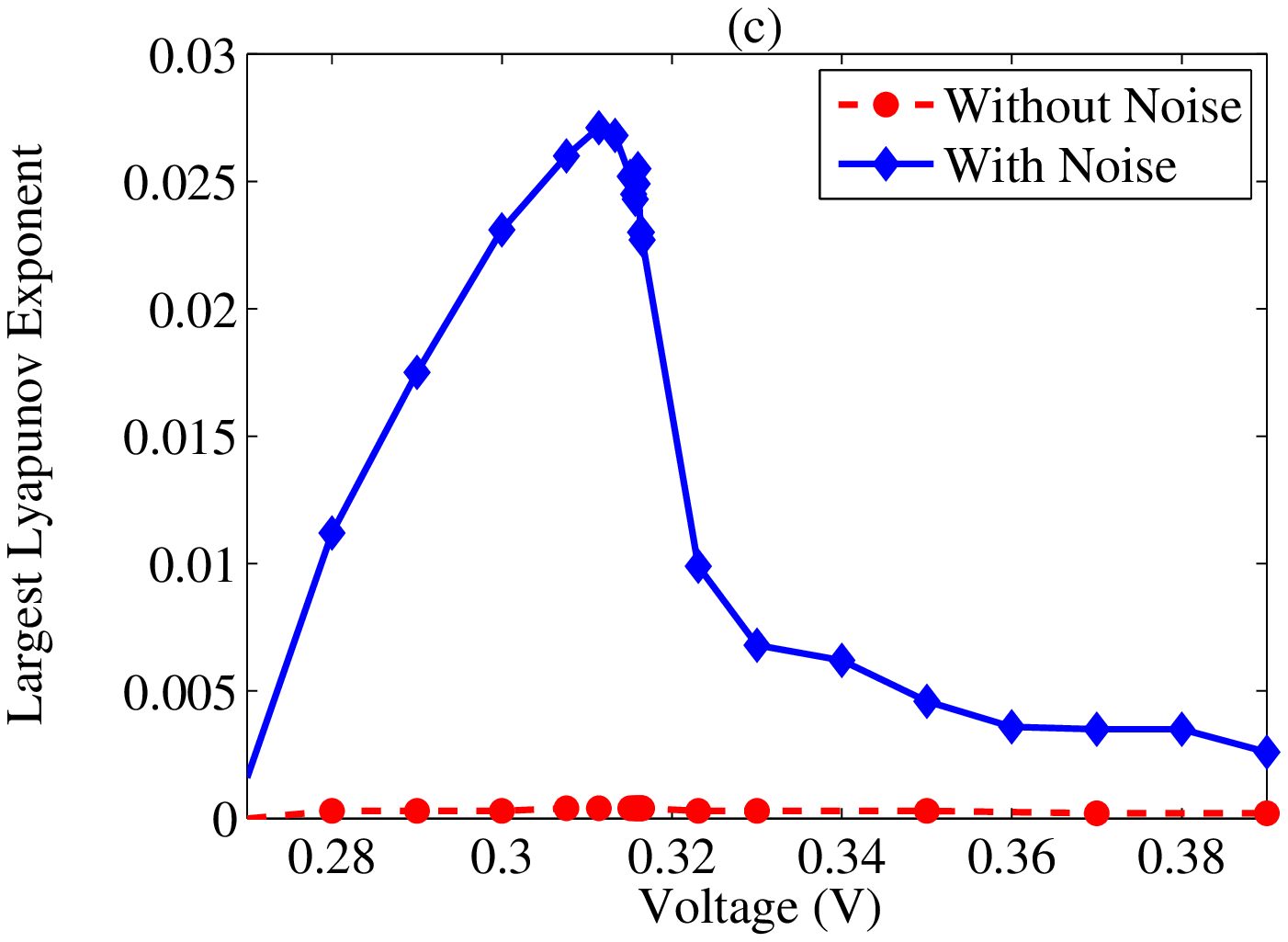}
\end{center}
\caption{(a) Current response to voltage upsweep and then downsweep of the deterministic system showing that the self-oscillations start as a subcritical Hopf bifurcation from the stationary state. (b) Fourier spectrum of the current vs voltage for zero noise. (c) Largest Lyapunov exponent vs voltage.  The LLE curves have the same shape for the deterministic and stochastic cases which cannot be appreciated due to their very different scales: Noise increases LLEs from 0.0004 up to 0.028.
} \label{fig2} 
\end{figure}

\begin{figure}[htbp]
\begin{center}
\includegraphics[width=8cm]{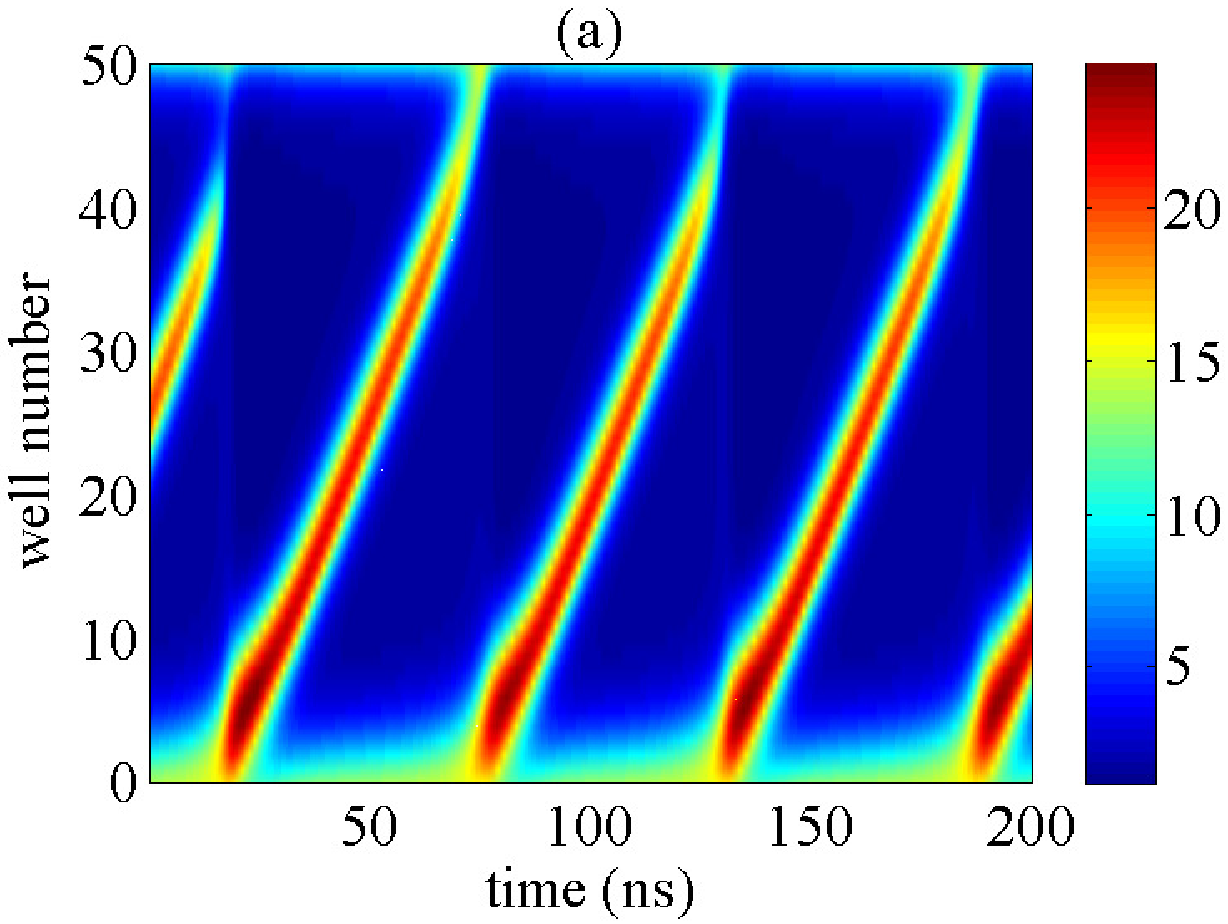}\\
\includegraphics[width=8cm]{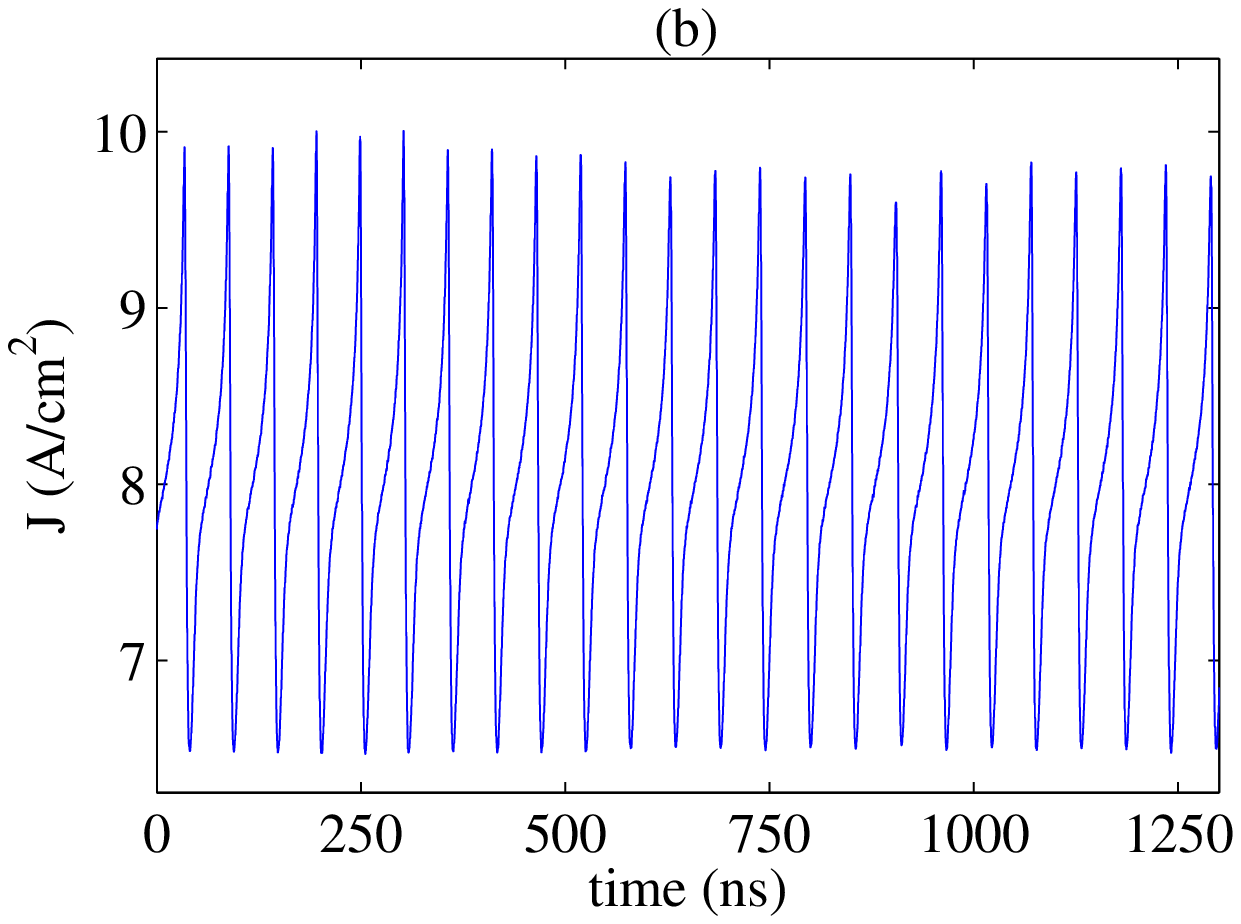}\\
\includegraphics[width=8cm]{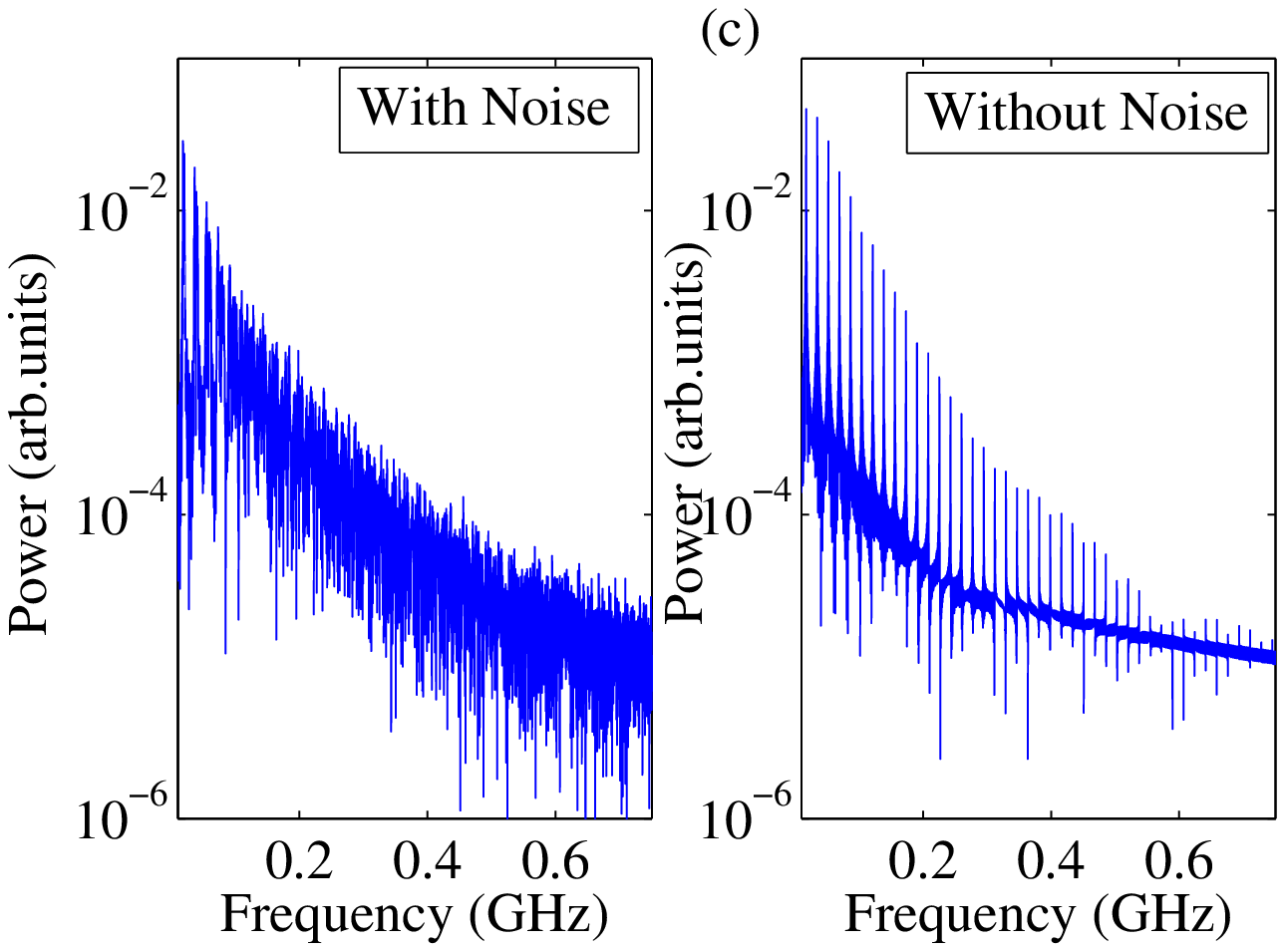}
\end{center}
\caption{(a) Density plot of the electric field vs time, and (b) current  vs time, showing how the dipole occasionally fails to reach the receiving contact. (c) Fourier transform of the current trace. The superlattice is biased at 0.315 V. 
} \label{fig3} 
\end{figure}

We have solved the stochastic model given by Eqs.~(\ref{eq1})-(\ref{eq7}) for the SL of Refs.~\cite{li13,hua12,hua13} at 300 K using a standard stochastic Euler method. To calculate the largest Lyapunov exponent (LLE), we have simultaneously integrated all perturbed and unperturbed trajectories and used the Benettin {\em et al} algorithm \cite{ben78}. LLE calculations with the Gao {\em et al} algorithm \cite{gao98,gao94} give the same results. The deterministic (noiseless) system exhibits self-sustained oscillations in two voltage intervals or plateaus. In the first plateau (corresponding to the one reported in \cite{li13}), large-amplitude current oscillations appear as a subcritical Hopf bifurcation from the stationary state, as shown in Fig.~\ref{fig2}(a). They are caused by the repeated creation of field pulses (charge dipoles) that dissolve before arriving at the collector. For larger voltages, there is a transition from large amplitude and frequency to smaller amplitude and frequency current oscillations. The transition to a two-frequency oscillation of richer harmonic content is clearly observable in the Fourier spectrum of Figure~\ref{fig2}(b). In the transition region, a new frequency appears and Figure~\ref{fig2}(c) shows that the LLE becomes positive indicating sensitivity to initial conditions and chaos. Direct routes to chaos from two-frequency quasi-periodic attractors have been shown in simpler dynamical systems \cite{moon}. Figure~\ref{fig2}(c) shows that the deterministic system is weakly chaotic in a narrow voltage interval and that internal and external noises enhance chaos for these voltages. For larger voltages the oscillation again has a single dominant frequency. For the stochastic system, Fig.~\ref{fig3}(a) and (b) display the electric field profile and the current traces vs time at a voltage for which the LLE is positive. In addition, noise enriches the content of the power spectrum as shown by Fig.~\ref{fig3}(c). That two oscillatory modes are present in the observed spontaneous chaotic oscillations is commented in Ref.~\cite{hua13}, whose authors identify them as the dipole motion mode and the well-to-well hopping mode. In Fig.~\ref{fig3}(a), they should correspond to dipoles reaching the collector in their motion and to the confined dipole motion, respectively. 

\begin{figure}[htbp]
\begin{center}
\end{center}
\includegraphics[width=8cm]{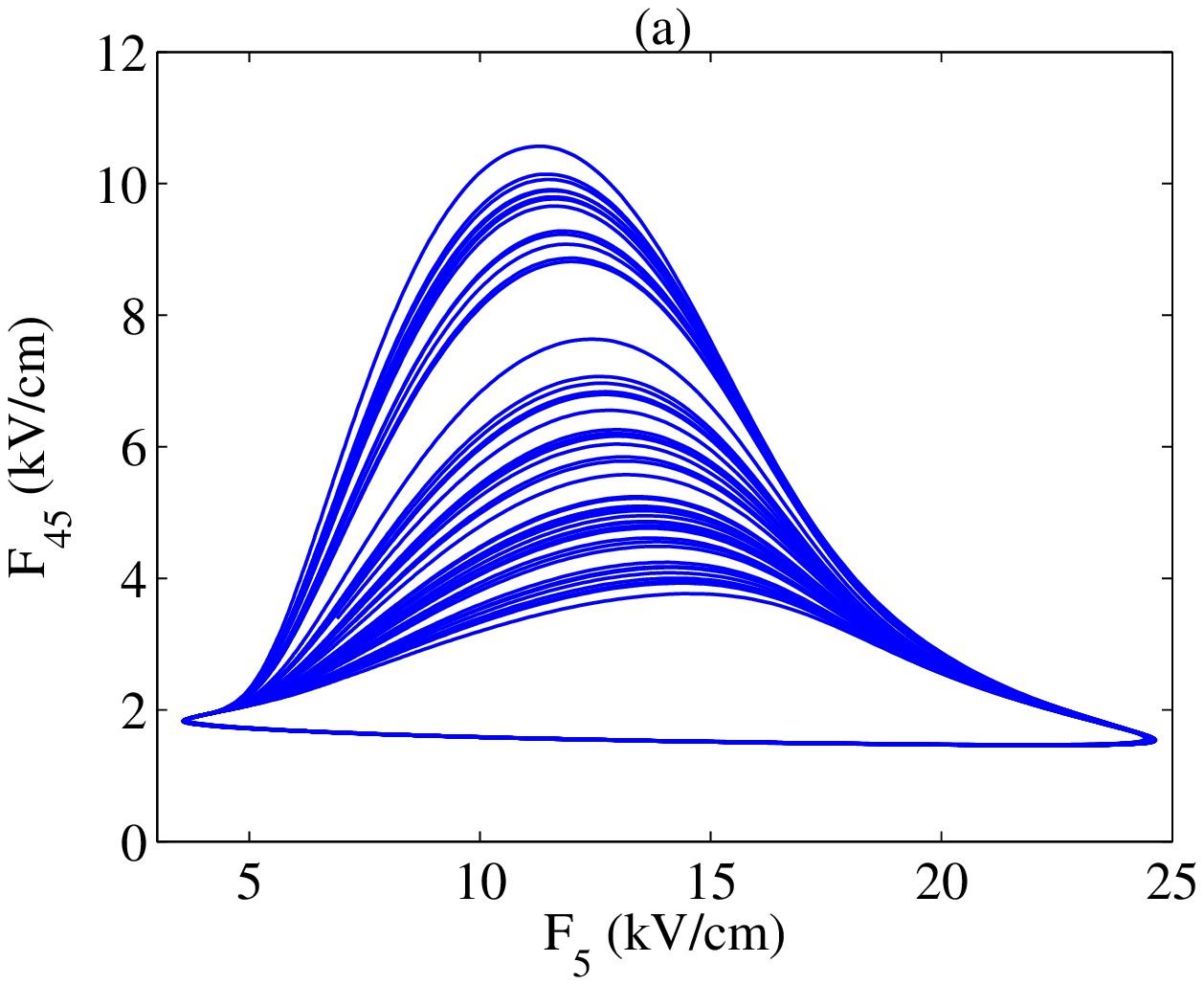}\\
\includegraphics[width=8cm]{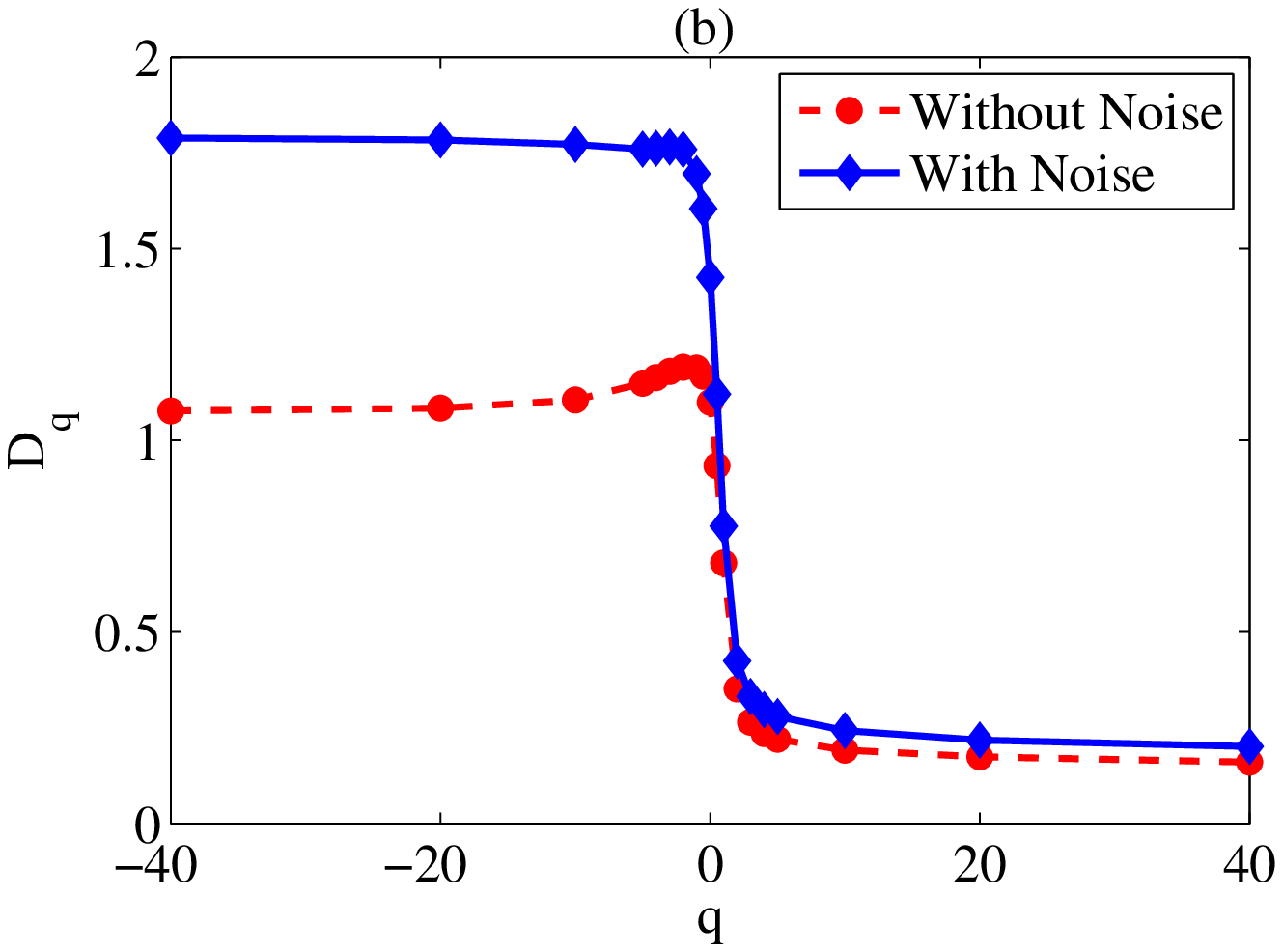}
\caption{(a) Phase diagram of the electric field at wells 5 and 45, showing how the chaotic oscillations fill a snail-shaped region; (b) multifractal dimension. Applied voltage is 0.315 V.}
\label{fig4}
\end{figure}

Internal and external noises affect nonlinear charge transport in SLs in ways that are currently being explored. Noise-induced current switching between stable stationary states has been reported in a weakly coupled GaAs/AlAs SL that acts as an excitable system \cite{bom12}. Noise induced coherence resonance predicted in \cite{hiz06} has been observed in another GaAs/AlAs SL at 77 K \cite{hua14}. Here we propose the noise-enhanced-chaos mechanism to explain for the first time spontaneous chaotic oscillations in a SL at room temperature and confirm it by numerical simulation. Essentially, the noise in a SL may convert regions where the deterministic description of the system presents sharp transitions between different oscillation modes into a chaotic attractor. The chaotic attractor may exist in a narrow region of the deterministic system, but the noise then enforces and changes it into the chaotic attractor of higher fractal dimension shown in Figures~\ref{fig4}(a) and (b). (The multifractal dimension has been calculated using the methods explained in Chapter 9 of Ref.~\cite{ott} and in Refs.~\cite{paw87,bul99}). We have checked that noise enhances chaos if its level falls within a narrow range, as indicated in Ref.~\cite{gao98} for simpler systems. Too small a noise does not induce chaos whereas too large a noise swamps spontaneous chaos. Far from the transition, each oscillatory mode is globally stable, and the noise cannot induce an attractor with extreme sensitivity to initial conditions quantified by a positive LLE. 

At the voltage corresponding to the maximum value of the LLE in Fig.~\ref{fig2}(c), the fluctuations create a snail-shaped {\em multifractal} chaotic attractor out of the above mentioned small and large amplitude oscillatory modes, as shown in Fig.~\ref{fig4}(a). The current trace shows several irregularly separated large and smaller spikes that indicate nucleation of a new dipole at the injector. Large (small) spikes occur when a dipole disappears before (after) reaching the collector. When a dipole dies before reaching the collector, the corresponding electric field at the injector (and, by (\ref{eq6}), the total current) is larger than that corresponding to a dipole that reaches the collector. Thus, two different oscillation modes are observable in the chaotic attractor: injector-to-collector dipole motion and dipole motion from injector to premature annihilation inside the SL. The latter corresponds to the well-to-well hopping mode postulated in Ref.~\cite{hua13}. The inter-spike intervals are similar but never repeat themselves and tend to produce a snail-shaped attractor in the phase plane of the field at two separated wells as depicted in Fig.~\ref{fig4}(a). 

We have predicted that spontaneous chaos appears at room temperature dc voltage biased SLs and is strongly enhanced by internal and external noises. Time periodic current oscillations appear as subcritical Hopf bifurcations for voltages on the first plateau and we have checked that they do occur as supercritical Hopf bifurcations on the second plateau. In both plateaus, a second oscillation frequency appears after the Hopf bifurcations and spontaneous chaos occurs near the corresponding transition voltage. Unavoidable internal and external noises enhance chaos: LLE and multifractal dimension of attractors increase with respect to the deterministic case. It is known that adding  external noise may suppress or enhance the effects of internal noise on a given system \cite{vil01}. In our system, increasing the complexity of the chaotic attractor by adding controlled sources of external noise may speed up random number generation. This is what we find for an ideal model of perfect SLs with sequential tunneling as their only vertical electron transport mechanism. 

Our results differ from experimental reports \cite{li13} in two aspects. Firstly, the voltage intervals for spontaneous chaos seem wider in experiments (0.3 V instead of 0.13 V) and the oscillations are more irregular than ours. Secondly, the oscillation frequencies are about 8 times larger in the experiments and the mean current density is about 26 times larger (260 instead of 10 A/cm$^2$). However, the first plateau is located at comparable voltages in theory and experiments once we subtract the voltage at the series resistance (3V in the experiments, negligible in the model). Both discrepancies have to do with the limitations of our model. Real samples have fluctuations in the aluminium density at the barriers, the width of barriers and wells and the well doping density.  These effects could be included in our model by adding time-independent but well-dependent random variables to the tunneling current and to the doping density. This means that the functional form of the tunneling current density may vary from SL well to well. This in turn may alter and enrich charge dipole dynamics responsible for chaos in the real samples but exploring this is outside the scope of this paper. About the discrepancy in the frequency scales, the only mechanism for electron transport included in the model is sequential resonant tunneling. This means that the current and the frequency are too small compared to experimental values as already reported in \cite{rog01}. 

In conclusion, we have explained for the first time spontaneous chaos in semiconductor superlattices at room temperature as being associated to sharp transitions between oscillatory modes. Internal and external noises enhance spontaneous chaos in these devices. Increasing the complexity of the chaotic attractor by adding controlled sources of external noise may speed up random number generation. Our work paves the way for the design and optimization of superlattice-based devices that exploit spontaneous chaos to generate truly random numbers at high speed \cite{li13}. Such random number generators are of paramount importance in encryption systems for data storage and transmission \cite{sti95,gal08,nie00}, stochastic modeling \cite{asm07}, and Monte Carlo simulation \cite{bin02}.

\acknowledgments This work has been supported by the Spanish Ministerio de Econom\'\i a y Competitividad grant FIS2011-28838-C02-01.

\appendix
\section{Internal noise}
We have considered that the internal noise is due to shot and thermal noise \cite{bla00}. Shot/partition noise for our discrete model of SL transport is due to charge quantisation when electrons cross SL barriers \cite{bla00}, as indicated in Ref.~\cite{bon04}, and in Ref.~\cite{bon02} for a somewhat simpler discrete transport model. We consider current fluctuations associated to dissipation due to electron diffusion and model them by Landau-Lifshitz fluctuating hydrodynamics \cite{ll6,kei87} adapted to SLs. In this way, our thermal noise is based on general principles of local equilibrium and detailed balance and we avoid more specific noise modeling. Current fluctuations are due to electron diffusion in the discrete model of Eqs.~(\ref{eq1})-(\ref{eq3}). The latter can be written as a discrete drift-diffusion current density,
\begin{eqnarray}
J_{i\to i+1}\!\!\!&=&\!\!\frac{en_i}{l}v^{(f)}(F_i)-J_{i\to i+1}^-(F_i,n_{i},T)\nonumber\\
\!\!\!&-&\!\! [J_{i\to i+1}^-(F_i,n_{i+1},T)-J_{i\to i+1}^-(F_i,n_{i},T)], \label{eq8}
\end{eqnarray}
where the last two terms correspond to electron diffusion. According to fluctuating hydrodynamics \cite{ll6,kei87}, the current density fluctuations are zero-mean white noises with correlation 
$\frac{2e}{A}J_{i\to i+1}^-(F_i,n_{i},T)$, proportional to the diffusion current in Eq.~(\ref{eq4}), \cite{ll6,kei87}. This is similar to fluctuations in Gunn diodes where the fluctuation of the diffusion current is proportional to the diffusion coefficient times the electron density \cite{kei87} instead of the expression (\ref{eq5}). 

\section{Numerical issues}
To solve the Ito stochastic differential equations of the model (\ref{eq1})-(\ref{eq7}), we have used a standard stochastic Euler-Maruyama method \cite{kloeden} with 100 fs time steps. To calculate the LLE, we take an arbitrary  initial perturbation vector $d_0$ and simultaneously integrate both perturbed and unperturbed trajectories during 10,000 ns. The LLE is given by:
\begin{eqnarray*}
\lambda = \lim_{t \to \infty} \frac{1}{t} \ln \frac{\|d(t)\|}{\|d_0\|}
\end{eqnarray*}
where $\|d(t)\|$ is the distance in phase space between the perturbed and unperturbed trajectories.
The perturbation will need to be renormalized from time to time in order to prevent accumulative
numerical rounding errors. To this end, we use the Benettin {\em et al} algorithm \cite{ben78}
with a renormalization period of $\tau_r=1$ ns. Then 
\begin{eqnarray*}
\lambda = \lim_{k \to \infty} \frac{1}{k \tau_r} \sum_{j=1}^k \ln \frac{\|d_j\|}{\|d_0\|}
\end{eqnarray*}
where $d_j$ is the perturbation vector during the $j$th renormalization period.

We have also calculated the LLE by the  quite different Gao {\em et al} algorithm
\cite{gao94,gao98}. In this method, numerical simulation yields the current $J(t)$ sampled every
$\delta t \approx 1$ ns, which produces 4,000 values of the current. Then we calculate
$m$-dimensional vectors $J_i=(J(t_i),\ldots,J(t_i-(m-1)L\delta t))$, where $m$ is the smallest
integer equal or larger than twice the box-counting fractal dimension and, for different $k$, the
average
\begin{eqnarray*}
\Lambda(k)=\left\langle\ln\frac{\|J_{i+k}-J_{j+k}\|}{\|J_i-J_j\|}\right\rangle\!,\label{eq9}
\end{eqnarray*}
with $\|J_i-J_j\| < r$. The function $\Lambda(k)$ is linear over an extended interval and its slope
(written in nondimensional units) is the sought LLE, which coincides with that obtained by the
Benettin {\em et al} algorithm.

\end{document}